\documentclass[aps,prl,twocolumn,showpacs]{revtex4}
\usepackage[]{graphicx}
\usepackage{bm}
\usepackage{amsmath}

\begin{document}

\title{Topological phase for spin-orbit transformations on a laser beam}
 
\author{C. E. R. Souza$^{\dagger}$, J. A. O. Huguenin$^{\dagger}$, 
P. Milman$^{\dagger\dagger}$, and A.Z. Khoury$^{\dagger}$}

\affiliation{$^{\dagger}$
Instituto de F\'\i sica, Universidade Federal Fluminense,
24210-340 Niter\'oi - RJ, Brasil.}
\affiliation{$^{\dagger\dagger}$
Laboratoire Mat\'eriaux et Phenom\`enes quantiques CNRS UMR 7162,
Universit\'e Denis Diderot, 2 Place Jussieu 75005 Paris cedex.}

 
\begin{abstract}
We investigate the topological phase associated with the double connectedness 
of the $SO(3)$ representation in terms of maximally entangled states. An 
experimental demonstration is provided in the context of polarization and spatial 
mode transformations of a laser beam carrying orbital angular momentum. 
The topological phase is evidenced through interferometric measurements and 
a quantitative relationship between the concurrence and the fringes visibility 
is derived. Both the quantum and the classical regimes were investigated. 
\end{abstract}
\vskip2pc 
 
\maketitle

The seminal work by S. Pancharatnam \cite{pancha} introduced for the first time 
the notion of a geometric phase acquired by an optical beam passing through 
a cyclic sequence of polarization transformations. A quantum mechanical 
parallel for this phase was later provided by M. Berry \cite{berry}. 
Recently, the interest for geometric phases was renewed by their potential 
applications to quantum computation. The experimental demonstration of 
a conditional phase gate was recently provided both in nuclear magnetic 
ressonance \cite{vedral} and trapped ions \cite{zoller}.
Another optical manifestation of geometric phase is the one acquired by 
cyclic spatial mode conversions of optical vortices. This kind of geometric 
phase was first proposed by van Enk \cite{vanenk} and recently found a 
beautiful demonstration by E. J. Galvez {\it et al} \cite{galvez}.

The Hilbert space of a single qubit admits an useful geometric representation of 
pure states on the surface of a sphere.  This is the Bloch sphere for 
spin $1/2$ particles or the Poincar\'e sphere for polarization states of 
an optical beam. 
A Poincar\'e sphere representation can also be constructed for the first 
order subspace of the spatial mode structure of an optical beam \cite{padgett}. 
Therefore, in the quantum domain, we can attribute 
two qubits to a single photon, one related to its 
polarization state and another one to its spatial structure. 
Geometrical phases of a cyclic evolution of the mentioned states can
be beautifully interpreted in such representations as being related to
the solid angle of a closed trajectory. However, in order to compute
the total phase gained in a cyclic evolution, one should also consider
the dynamical phase. When added to the geometrical phase, it leads to
a total phase gain of $\pi$ after a cyclic trajectory. This phase has
been put into evidence for the first time using neutron interference
\cite{neutrons}.
The appearence of this $\pi$ phase is due to the double connectedness
of the three dimensional rotation group $SO(3)$. However, in the
neutron experience, only two dimensional rotations were used, and this
topological property of $SO(3)$ was not unambiguously put into
evidence, as explained in details in \cite{remy,remy2}. 

As discussed by 
P. Milman and R. Mosseri \cite{remy,milman}, when the quantum state of two 
qubits is considered, the mathematical structure of the Hilbert space becomes 
richer and the phase acquired through cyclic evolutions demands a more careful 
inspection. The naive sum of independent phases, one for each qubit, is 
applicable only for product states. In this case, the two qubits are 
geometrically represented by two independent Bloch spheres.
When a more general partially entangled 
{\it pure} state is considered, the phase acquired through a cyclic evolution 
has a more complex structure and can be separated in three contributions: 
dynamical, geometrical and topological. 
Maximally entangled states are solely represented on 
the volume of the $SO(3)$ sphere which has radius $\pi$ and its diametrically 
opposite points identified. This construction reveals two kinds of cyclic 
evolutions, each one mapped to a different homotopy class of closed trajectories 
in the $SO(3)$ sphere. One kind is mapped to closed trajectories that do not 
cross the surface of the sphere ($0-$type) and the other one is mapped to 
trajectories that cross the surface ($\pi-$type). 
The phase acquired by a maximally entangled state is $0$ for the first kind 
and $\pi$ for the second one. 

In the present work we demonstrate the topological phase associated to 
polarization and spatial mode transformations of an optical vortice. This 
phase appears first in the classical description of a paraxial beam with 
arbitrary polarization state and has its quantum mechanical counterpart 
in the spin-orbit entanglement of a single photon, which constitutes one 
possible realization of a two-qubit system and the topological phase 
discussed in Ref.\cite{remy}. However, it is interesting to 
observe that, like the Pancharatnam phase, the two-qubit topological 
phase also admits a classical manifestation, since it can be implemented 
on the classical amplitude of the optical field.
This is also the first experiment
unambiguously showing the double connectedness of the rotation
group $SO(3)$. The optical modes used in our experiment have a
mathematical structure analog to the one of entangled states, so that
the geometrical representation developped in \cite{remy2} also applies 
and the results of Ref.\cite{remy,milman} can be experimentally 
demonstrated. When excited with single photons, these modes give 
rise to single particle entangled states and provide a more direct 
relationship with the ideas put forward in 
Refs.\cite{remy,remy2,milman}. This regime is also investigated in the 
present work. There are a number of quantum computing 
protocols that can be implemented with single particle entanglement and 
will certainly benefit from our results.

Let us now 
combine the spin and orbital degrees of freedom in the framework of the 
classical theory in order to build the same geometric representation 
applicable to a two-qubit quantum state. Consider a general first order 
spatial mode with arbitrary polarization state:
\begin{eqnarray}
\mathbf{E}(\mathbf{r})&=&
\alpha \psi_+(\mathbf{r})\hat{e}_H +
\beta \psi_+(\mathbf{r})\hat{e}_V +
\gamma \psi_-(\mathbf{r})\hat{e}_H 
\nonumber\\
&+& \delta \psi_-(\mathbf{r})\hat{e}_V\;,
\label{genmode}
\end{eqnarray}
where $\hat{e}_{H(V)}$ are two linear polarization unit vectors along two 
orthogonal directions $H$ and $V$, and $\psi_{\pm}(\mathbf{r})$ are the 
normalized first order Laguerre-Gaussian profiles which are orthogonal 
solutions of the paraxial wave equation \cite{yariv}. We may now define 
two classes of spatial-polarization modes: the separable (S) and the 
nonseparable (NS) ones. The S modes are of the form
\begin{equation}
\mathbf{E}(\mathbf{r})=
\left(\alpha_+ \psi_+(\mathbf{r}) + 
\alpha_- \psi_+(\mathbf{r})\right)
\left(\beta_H \hat{e}_H +
\beta_V\hat{e}_V\right)\;.
\label{smode}
\end{equation}
For these modes, a single polarization state can be atributted to the 
whole wavefront of the paraxial beam. They play the role of separable 
two-qubit quantum states. 

For nonseparable (NS) paraxial modes, the polarization state varies 
across the wavefront. As for entanglement in two-qubit quantum states, 
the separability of a paraxial mode can be quantified by the analogous 
definition of concurrence. For the spin-orbit mode described by 
Eq.(\ref{genmode}), it is given by:
\begin{equation}
C=2\mid \alpha\delta-\beta\gamma\mid\;.
\label{conc}
\end{equation}

Let us first consider the maximally nonseparable modes (MNS) of the form
\begin{eqnarray}
\mathbf{E}(\mathbf{r})&=&
\alpha \psi_+(\mathbf{r})\hat{e}_H +
\beta \psi_+(\mathbf{r})\hat{e}_V -
\beta^* \psi_-(\mathbf{r})\hat{e}_H 
\nonumber\\
&+& \alpha^* \psi_-(\mathbf{r})\hat{e}_V\;.
\label{mnsmode}
\end{eqnarray}
For these modes $C=1$. 
It is important to mention that the concept of entanglement 
does not applies to the MNS mode, since the object described by 
Eq.(\ref{mnsmode}) is not a quantum state, but a classical amplitude. 
However, we can build an $SO(3)$ representation of the MNS modes as it was done 
in Refs.\cite{milman,chines}. Let us define the following normalized 
MNS modes:
\begin{eqnarray}
\mathbf{E_1}(\mathbf{r})=\frac{1}{\sqrt{2}}
\left[
\psi_+(\mathbf{r})\hat{e}_H +
\psi_-(\mathbf{r})\hat{e}_V
\right]\;,
\nonumber\\
\mathbf{E_2}(\mathbf{r})=\frac{-i}{\sqrt{2}}
\left[
\psi_+(\mathbf{r})\hat{e}_H -
\psi_-(\mathbf{r})\hat{e}_V
\right]\;,
\label{modes1234}\\
\mathbf{E_3}(\mathbf{r})=\frac{-i}{\sqrt{2}}
\left[
\psi_+(\mathbf{r})\hat{e}_V +
\psi_-(\mathbf{r})\hat{e}_H
\right]\;,
\nonumber\\
\mathbf{E_4}(\mathbf{r})=\frac{1}{\sqrt{2}}
\left[
\psi_+(\mathbf{r})\hat{e}_V -
\psi_-(\mathbf{r})\hat{e}_H
\right]\;.
\nonumber
\end{eqnarray}
The $SO(3)$ sphere is then constructed in the following way: mode 
$\mathbf{E_1}$ is represented by the center of the sphere, while modes 
$\mathbf{E_2}$, $\mathbf{E_3}$, and $\mathbf{E_4}$ are represented by 
three points on the surface, connected to the center by three mutually 
orthogonal segments.
Each point of the $SO(3)$ sphere corresponds to a MNS mode. Following the 
recipe given in Ref.\cite{chines}, the coefficients $\alpha$ and $\beta$ 
of Eq.(\ref{mnsmode}) are parametrized to:
\begin{eqnarray}
\alpha&=&\cos\frac{a}{2}-i\,k_z\,\sin\frac{a}{2}\;,
\nonumber\\
\beta&=&-(k_y+i\,k_x)\,\sin\frac{a}{2}\;,
\end{eqnarray} 
where $(k_x,k_y,k_z)=\mathbf{k}$ is a unit vector, and $a$ is an angle 
between $0$ and $\pi$. With this parametrization, each MNS mode is 
represented by the vector $a\,\mathbf{k}$ in the sphere.

%
\begin{figure}
\begin{center} 
\includegraphics[scale=.5]{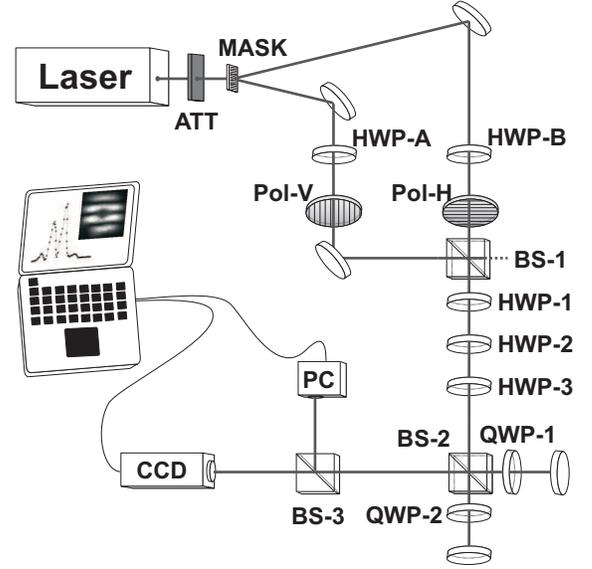}
\end{center} 
\caption{Experimental setup.}
\label{setup}
\end{figure}
%

In order to evidence the topological phase for cyclic transformations, 
we must follow two different closed paths, each one belonging to a 
different homotopy class, and compare their phases. The experimental 
setup is sketched in Fig.(\ref{setup}). First, a linearly polarized 
TEM$_{00}$ laser mode is diffracted on a forked grating used to generate 
Laguerre-Gaussian beams \cite{masks}. 
The two side orders carrying the 
$\psi_{+}(\mathbf{r})$ and $\psi_{-}(\mathbf{r})$ spatial modes are 
transmitted through half waveplates HWP-A and HWP-B, followed by 
two orthogonal polarizers Pol-V and Pol-H, and finally recombined 
at a beam splitter (BS-1). Half waveplates HWP-A and 
HWP-B are oriented so that their fast axis are paralell. This allows 
us to adjust the mode separability at the output of BS-1 without 
changing the corresponding output power, what prevents normalization 
issues. 

Experimentally, an MNS mode is produced when both HWP-A and HWP-B are 
oriented at $22.5^{o}\,$, so that the setup 
prepares mode $\mathbf{E_1}$ located at the centre of the sphere. Other 
MNS modes can then be obtained by unitary transformations in only one degree 
of freedom. Since polarization is far easier to operate than spatial modes we 
choose to implement the cyclic transformations in the $SO(3)$ sphere using 
waveplates. 
The MNS mode $\mathbf{E_1}$ is first transmitted through three waveplates. 
The first one (HWP-1) is oriented at $0^o$ and makes the transformation 
$\mathbf{E_1}\rightarrow\mathbf{E_2}$, the second one (HWP-2) is oriented at 
$-45^o$ and makes the transformation $\mathbf{E_2}\rightarrow\mathbf{E_3}$, 
and the third one (HWP-3) is oriented at $90^o$ and makes the transformation 
$\mathbf{E_3}\rightarrow\mathbf{E_4}$. Finally, two alternative closures 
of the path are performed in a Michelson interferometer. In one arm a 
$\pi-$type closure is implemented by double pass through a 
quarter-waveplate (QWP-1) fixed at $-45^o$. In the other arm, either a 
$0-$type or a $\pi-$type closure is performed by a double pass 
through another quarter-waveplate (QWP-2) oriented at a variable angle 
between $-45^o$ ($\pi-$type) and $45^o$ ($0-$type). 
These trajectories are analogous to spin rotations around 
different directions of space \cite{chines}. They evidence the 
topological properties of the three dimensional rotation group.

In order to provide spatial interference fringes, the interferometer 
was slightly misaligned. The interference patterns were registered 
with either a charge coupled device (CCD) camera or a photocounter 
(PC), depending on the working power.
%
\begin{figure}
\begin{center} 
\includegraphics[scale=.6]{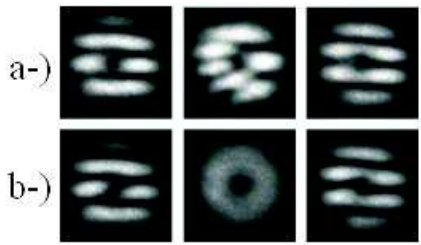}
\end{center} 
\caption{Interference patterns for a-) a maximally nonseparable, 
and b-) a separable mode. From left to right the images were obtained 
with QWP-2 oriented at $-45^o$, $0^o$, and $45^o$, respectively.}
\label{patterns}
\end{figure}
%
First, we registered 
the interference patterns obtained when an intense beam is sent through 
the apparatus. The images shown in Fig.(\ref{patterns}a) demonstrate clearly 
the $\pi$ topological phase shift. The phase singularity characteristic 
of Laguerre-Gaussian beams can be easily identified in the images and 
is very useful to evidence the phase shift. When both arms perform the 
same kind of trajectory in the $SO(3)$ sphere (QWP-1 and QWP-2 oriented 
at $-45^o$), a bright fringe falls on the phase singularity. When QWP-2 is 
oriented at $45^o$, the trajectory performed in each arm belongs to a 
different homotopy class and a dark fringe falls on the singularity, 
what clearly demonstrates the $\pi$ topological phase shift. 

In order to discuss the role played by mode separability, it is 
interesting to observe the pattern obtained when QWP-2 is oriented at 
intermediate angles, which correpond to open trajectories in the $SO(3)$ 
sphere. We observed that during the phase shift transition, the 
interference fringes are deformed and finally return to 
its initial topology with the $\pi$ phase shift. This is clearly 
illustrated by the intermediate image displayed in Fig.(\ref{patterns}a), 
which corresponds to QWP-2 oriented at $0^o\,$. Notice that, despite 
the deformation, the interference fringes display high visibility.

As we mentioned above, 
the mode preparation settings can be adjusted in order to provide a 
separable mode. For example, when we set HWP-A and B both at $45^o\,$, 
the output of BS-1 is the separable mode 
$\psi_+(\mathbf{r})\hat{e}_H\,$, which can be represented in the 
Poincar\'e spheres for spatial and polarization modes. 
The same $\pi$ phase shift can be observed when QWP-2 is rotated, but 
the transition is essentially different. The intereference pattern is 
not topologically deformed, but its visibility decreases until it 
completelly vanishes at $0^o\,$, and then reappears with the $\pi$ 
phase shift. This transition is clearly illustrated by the three 
patterns displayed in Fig.(\ref{patterns}b). In this case, the 
$\pi$ phase shift is of purely geometric nature, since the spatial 
mode is kept fixed while the polarization mode is turned around the 
equator of the corresponding Poincar\'e sphere.

The relationship between mode separability and fringes visibility can 
be clarified by a straightforward calculation of the interference 
pattern. Therefore, let us consider that HWP-A and B are oriented 
so that the output of BS-1 is described by 
\begin{equation}
\mathbf{E}_{\epsilon}(\mathbf{r})=\sqrt{\epsilon}\,
\psi_+(\mathbf{r})\hat{e}_H +
\sqrt{1-\epsilon}\,
\psi_-(\mathbf{r})\hat{e}_V\;,
\label{Eepsilon}
\end{equation}
where $\epsilon$ is the fraction of the $\psi_+(\mathbf{r})\hat{e}_H$ 
mode in the output power. Now, let us consider that QWP-2 is oriented 
at $0^o$ and suppose that the two arms of the Michelson interferometer 
are slightly misaligned so that the wave vectors difference between 
the two outputs is $\delta\mathbf{k}=\delta k\,\hat{x}\,$, orthogonal 
to the propagation axis. Taking 
into account the passage through the three half waveplates, and the 
transformation performed in each arm of the Michelson interferometer, 
we arrive at the following expression for the interference pattern: 
\begin{equation}
I(\mathbf{r})=2\,|\psi(\mathbf{r})|^2\left[
1+2\sqrt{\epsilon(1-\epsilon)}
\sin{2\phi}\sin{(\delta k\,x)}\right]\;,
\label{patterneq}
\end{equation}
where $\phi=\arg(x+iy)$ is the angular coordinate in the transverse 
plane of the laser beam, and $|\psi(\mathbf{r})|^2$ is the 
doughnut profile of the intensity distribution of a Laguerre-Gaussian 
beam. It is clear from Eq.(\ref{patterneq}) that the visibility of 
the interference pattern is $2\sqrt{\epsilon(1-\epsilon)}$, which 
is precisely the concurrence of $\mathbf{E}_{\epsilon}(\mathbf{r})$ 
as given by Eq.(\ref{conc}). Therefore, the fringes visibility is 
quantitatively related to the separability of the mode sent through 
the setup. However, the numerical coincidence with the concurrence 
is restricted to modes of the form given by Eq.(\ref{Eepsilon}). 
In fact, it is important to stress that the fringes visibility 
cannot be regarded as a measure of the concurrence for \textit{any} 
nonseparable mode, but for our purposes it evidences the topological 
nature of the phase shift implemented by the experimental setup. 
A detailed discussion on the measurement of the concurrence is 
available in Ref.\cite{concufrj}.

%
\begin{figure}
\begin{center} 
\includegraphics[scale=.7]{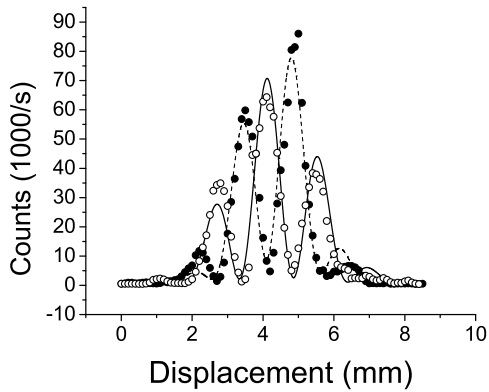}
\end{center} 
\caption{Interference patterns measured in the photocounting regime for 
$\epsilon = 1/2\,$. Empty and full circles correspond to QWP-2 oriented at 
$-45^o$ and $45^o$, respectively. Solid and dashed lines are theoretical 
fits with sinusoidal functions modulated by a Laguerre-Gaussian envelope. 
The phase shift given by the fits is $3.14\,rad\,$.}
\label{qpatterns}
\end{figure}
%

Next, we briefly discuss the quantum domain. When a partially  
nonseparable mode like $\mathbf{E}_{\epsilon}(\mathbf{r})$ is 
occupied by a single photon, this leads to partially entangled 
single particle quantum states of the kind
\begin{equation}
|\varphi_{\epsilon}\rangle = \sqrt{\epsilon}\,|+H\rangle + 
\sqrt{1-\epsilon}\,|-V\rangle\;.
\label{phiepsilon}
\end{equation}
Experimentally, we attenuated the laser beam down to the single photon 
regime, and scanned a photocounting module across the interference 
pattern. First, HWP-A and B were set at $22.5^o$ ($\epsilon = 1/2$) 
in order to evidence the topological phase in this regime. 
Fig.(\ref{qpatterns}) displays the interference patterns obtained 
with QWP-2 oriented at $-45^o$ and $45^o$. The $\pi$ phase shift is again 
clear. 

\begin{figure}
\begin{center} 
\includegraphics[scale=.7]{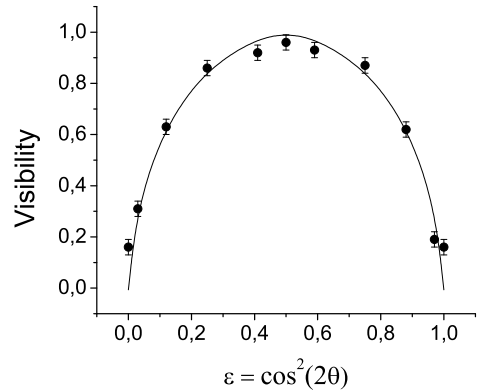}
\end{center} 
\caption{Fringes visibility as a function of $\epsilon$. The solid line is 
a theoretical fit with $C=2\sqrt{\epsilon(1-\epsilon)}\,$.}
\label{qconc}
\end{figure}

The relationship between the fringes visibility and the state separability 
was evidenced by fixing QWP-2 at $0^o$ and rotating HWP-A and B by an angle 
$\theta$ so that $\epsilon=\cos^22\theta\,$. Fig.(\ref{qconc}) 
shows the experimental  results for the fringes visibility for several values 
of $\epsilon\,$. The solid line corresponds to the analytical expression of 
the concurrence, showing a very good agreement with the experimental values. 
%
%

As a conclusion, we demonstrated the double connected nature of the
$SO(3)$ rotation group and the topological phase acquired by a laser 
beam passing through a cycle of spin-orbit transformations. We investigated 
both the classical and the quantum regimes and compared the separability 
of the mode travelling through the apparatus with the visibility of the 
interference fringes. Our results may constitute an useful tool for 
quantum computing and quantum information protocols.

\acknowledgments
The authors are deeply grateful to S.P. Walborn and P.H. Souto Ribeiro for 
their precious help with the photocounting system and for fruitful discussions.
Funding was provided by Coordena\c c\~{a}o de Aperfei\c coamento de 
Pessoal de N\'\i vel Superior (CAPES), Funda\c c\~{a}o de Amparo \`{a} 
Pesquisa do Estado do Rio de Janeiro (FAPERJ-BR), and Conselho Nacional 
de Desenvolvimento Cient\'{\i}fico e Tecnol\'ogico (CNPq).

\end{document}